\documentclass[10pt,twocolumn,letterpaper]{article}

\usepackage{cvpr}              

\usepackage{graphicx}
\usepackage{amsmath}
\usepackage{amssymb}
\usepackage{booktabs}
\usepackage[export]{adjustbox}

\usepackage[pagebackref,breaklinks,colorlinks]{hyperref}

\def\blind#1{#1}

\usepackage[capitalize]{cleveref}
\crefname{section}{Sec.}{Secs.}
\Crefname{section}{Section}{Sections}
\Crefname{table}{Table}{Tables}
\crefname{table}{Tab.}{Tabs.}

\def\cvprPaperID{*****} 
\def\confName{CVPR}
\def\confYear{2023}

\begin{document}

\title{The Effect of Counterfactuals on Reading Chest X-rays}

\author{
Joseph Paul Cohen$^{1,2,\text{A}}$
\and
Rupert Brooks$^{3}$
\and
Sovann En$^{3}$
\and
Evan Zucker$^{1,2}$
\and
Anuj Pareek$^{1,2}$
\and
Matthew P Lungren$^{1,4,5}$
\and
Akshay Chaudhari$^{1,2}$
\and
\small $^1$ Stanford University Center for Artificial Intelligence in Medicine \& Imaging,\\
\small $^2$ Stanford University Department of Radiology, 
\small $^3$ Nuance Communications, \\
\small $^4$ University of California San Francisco Department of Radiology and Biomedical Imaging, 
\small $^5$ Microsoft Corporation
\\
}
\maketitle

\begin{abstract}
This study evaluates the effect of counterfactual explanations on the interpretation of chest X-rays. We conduct a reader study with two radiologists assessing 240 chest X-ray predictions to rate their confidence that the model's prediction is correct using a 5 point scale. Half of the predictions are false positives. Each prediction is explained twice, once using traditional attribution methods and once with a counterfactual explanation. The overall results indicate that counterfactual explanations allow a radiologist to have more confidence in true positive predictions compared to traditional approaches (0.15$\pm$0.95 with p=0.01) with only a small increase in false positive predictions (0.04$\pm$1.06 with p=0.57). We observe the specific prediction tasks of Mass and Atelectasis appear to benefit the most compared to other tasks. 
\end{abstract}

\paragraph{Introduction}

\let\thefootnote\relax\footnotetext{$^\text{A}$Work done prior to joining Amazon}

We study how counterfactual explanations impact a radiologist's ability to interpret the prediction of chest X-rays (CXRs) by a neural network model compared to traditional attribution methods. Specifically we focus on the issue of false positive predictions and if this can be mitigated using counterfactuals. A counterfactual explanation is one where the image features utilized by the model for a positive prediction are replaced with benign features which yield a negative prediction. Models can provide incorrect predictions when presented with out-of-distribution data or if they have learned spurious correlations. Even with rigorous testing before deployment these issues can still exist for rare inputs. In order to ensure that models are right for the right reasons \cite{Ross2017rrr}, who better than a domain expect to make this determination. By allowing the physician to visually confirm what features were used to make a prediction they can spot inconsistencies with the pathology and not accept the prediction.

\paragraph{Protocol and Materials}

Three traditional methods are used. The method $input$ $gradients$ (referred to as $grad$) which computes the absolute gradient of the input with respect to the prediction made for all images of the positive class $|\frac{\partial \hat{y_1}}{\partial \mathbf{x}}|$. The method \textit{Guided Backprop} \cite{Springenberg2014GuidedBackprop} (referred to as \textit{guided}) tries to ignore gradients that cancel each other out by only backpropagating positive gradients. The method \textit{Integrated Gradients} \cite{Sundararajan2017integrated} (referred to as \textit{integrated}) works by integrating gradients between the input image $x_i$ and an all-zero baseline image. 

The counterfactual method used is the Latent Shift method \cite{Cohen2021gif}. This method utilizes an autoencoder to regularize the pixel changes in order to decrease the prediction of a classifier. This overcomes a common issue creating adversarial images where the changes are imperceptible. The method generates a short looping video showing the removal of the features used by the model to make the prediction. Each frame is a step in the transformation of removing the feature used by the classifier to make the prediction.

The classifier utilized is a DenseNet-121 (called XRV-all) from the TorchXRayVision library \cite{Cohen2020xrv} which is jointly trained on the 7 largest CXR datasets (NIH, PC, CheX, MIMIC-CXR, Google, OpenI, RSNA). The performance on the pathologies under test are all above 0.77 AUC.

\begin{figure*}[h!]
    \centering
\vspace{-5pt}
\includegraphics[width=0.53\textwidth,fbox]{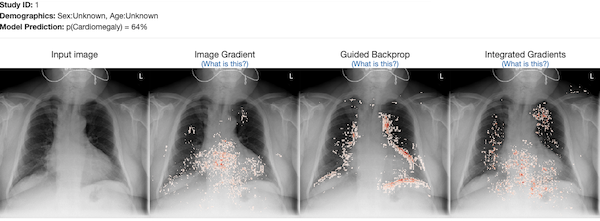} \includegraphics[width=0.43\textwidth,fbox]{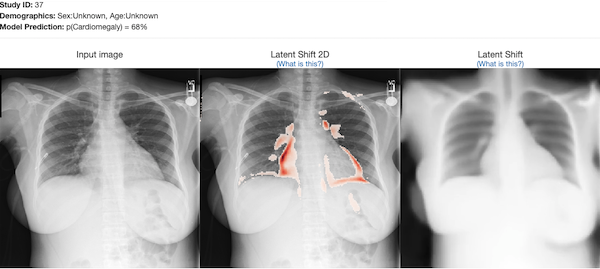}
\vspace{-5pt}
    \caption{\small Screenshots of the per image interface used in the reader study. On the left is the interface with Traditional methods and on the right is when using the counterfactual method. The counterfactual is shown as an animated image showing the transformation between the input image and the counterfactual version. A heatmap is provided to help the user identify the changes.}
    \label{fig:interfaceviewer}
\end{figure*}

\begin{figure*}[t]
    \centering
    \vspace{-5pt}
    \includegraphics[width=0.97\textwidth]{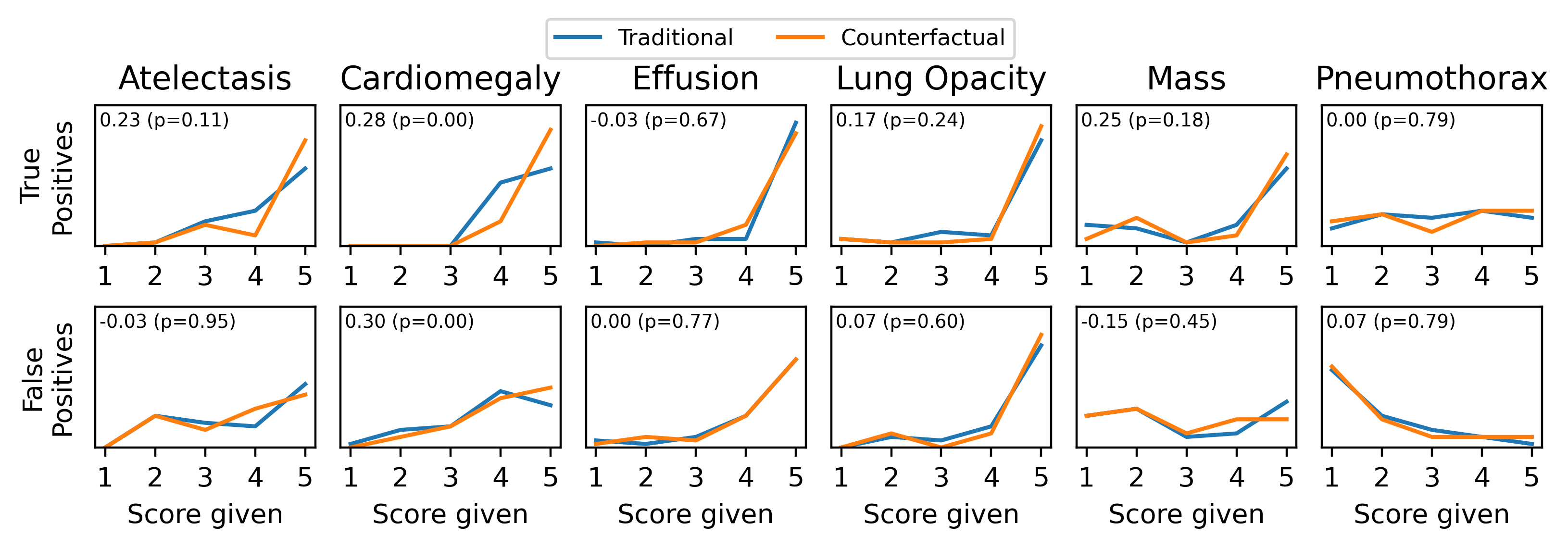}
    \vspace{-10pt}
    \caption{\small Histograms of responses to the survey question "How confident are you in the model's prediction?" split by each pathology. Plots are normalized to the maximum number of responses for each pathology (40). An optimal method would result in a score of 5 for all true positives and 1 for false positives. The difference between the methods (counterfactual-traditional) and a p-value is shown in the upper left of each plot. Overall the counterfactual method produced higher confidence for true positives and no significant impact on false positives.}
    \label{fig:reader-results}
\end{figure*}

\paragraph{Study Details}
For this study we recruited two radiologists (\blind{A.J. and E.Z.}, with 2 and 12 years of experience, respectively). They were presented with 240 images (from the NIH ChestXray14 dataset \cite{WangNIH2017}) twice, each being predicted as having at least one of 6 pathologies by the XRV-all model (Atelectasis, Cardiomegaly, Effusion, Lung Opacity, Mass, Pneumothorax). 40 images per pathology were selected such that 50\% were predicted incorrectly by the model (false positives). An incorrect prediction is defined by having a negative label and a $>$50\% prediction by the model which was calibrated such that a 50\% prediction is the operating point of the AUC curve on validation data. 

Each sample is viewed using traditional attribution methods and the counterfactual method using the interface shown in Figure \ref{fig:interfaceviewer}. For traditional methods the radiologist can see all methods at once (grad, guided, and integrated). For the counterfactual the radiologist can see an animated image showing the transformation between the input image and the counterfactual as well as a heatmap indicating where the changes are. Radiologists were asked the question on a 5 point Likert scale: ``How confident are you in the model's prediction? (1-5)".

The primary study results are shown in Figure \ref{fig:reader-results}. 
Overall, for true positive predictions there is a 0.15$\pm$0.95 confidence increase using the counterfactual method ($p$=0.01 using the Wilcoxon signed-rank test). Overall for false positive predictions there is a 0.04$\pm$1.06 increase which is not significant ($p$=0.57). 

The most notable improvements are observed for the Mass and Atelectasis prediction tasks. Predictions of Mass yielded a 0.25 confidence increase for true positives compared to a -0.15 confidence decrease for false positives. Predictions of Atelectasis yielded a 0.23 confidence increase for true positives compared to a -0.03 confidence decrease for false positives. These results are promising as these tasks have the largest change between true and false positives and this trend would indicate improvements in reading performance. One insight into this improvement is that the classifier would confuse a port with a Mass and the counterfactual would clearly indicate this by shrinking the size of the port making it clear the classifier did not see a Mass.

In the radiologist's feedback they believed that the counterfactual method was more intuitive and they felt it increased their confidence that the model is looking at the correct feature. They observed that this method looks at the boundaries of the abnormality. One radiologist believed that the model was using the chest tube, a spurious correlation, to predict Pneumothorax instead of looking at the correct area.

We find that counterfactual explanations allow a radiologist to have more confidence in true positive predictions compared to traditional approaches. However, we also found that detecting false positive predictions was challenging, which highlights the need for a stronger radiologist-algorithm symbiosis.

This extended abstract highlights an experiment performed in the work \cite{Cohen2021gif}. Please refer to this work for more details.

\section*{Ethics}
\noindent Only previously public patient data was used in this work. The study conducted in this research has been approved by the ethical review board at Stanford University.

{\small
\bibliographystyle{ieee_fullname}
\bibliography{cohen21}
}

\appendix
\counterwithin{figure}{section}
\counterwithin{table}{section}

\renewcommand\thefigure{\thesection.\arabic{figure}}

\end{document}